\documentstyle[12pt]{article}

\title{ A Comment on Jones Inclusions with infinite Index }
\author{{\sc Florian Nill,  Hans-Werner Wiesbrock
\thanks{EMAIL NILL@ omega.physik.fu-berlin.de\ ,  \ WIESBROC@
omega.physik.fu-berlin.de}}
\\Institut f\"ur Theoretische Physik \\ FU Berlin
\thanks{supported by the DFG, SFB 288 ``Differentialgeometrie und
Quantenphysik''}}

\newcommand{\be}{\begin{equation}}

\newcommand{\ee}{\end{equation}}

\newcommand{\bea}{\begin{eqnarray}}

\newcommand{\ea}{\end{eqnarray}}

\newcommand{\nn}{\nonumber}

\newcommand{\ca}{{\cal A}}

\newcommand{\cao}{\ca_{-1,0} }

\newcommand{\cae}{\ca_{-2,-1} }

\newcommand{\cm}{{\cal M}}

\newcommand{\cn}{{\cal N}}

\newcommand{\ch}{{\cal H}}

\newcommand{\cf}{{\cal F}}

\newcommand{\eot}{\hfill $\Box$\vspace{1cm}}

\newcommand{\jm}{J_{\cal M}}

\newcommand{\jn}{J_{\cal N}}

\newcommand{\lam}{\lambda}

\newcommand{\li}{\lambda_i}

\newcommand{\lj}{\lambda_j}

\newcommand{\om}{\Omega}

\newcommand{\proof}{\noindent {\bf Proof :}}

\newtheorem{theorem}{Theorem }

\newtheorem{proposition}[theorem]{Proposition }

\newtheorem{lemma}[theorem]{Lemma }

\newtheorem{corollary}[theorem]{Corollary }

\newtheorem{definition}[theorem]{Definition }

\begin{document}

\maketitle

$$ (\mbox{{\large dedicated to Bert Schroer's 60th birthday }}) $$

\begin{abstract}
Given an irreducible inclusion of infinite von-Neumann-algebras
$\cn \subset \cm$ together with a conditional expectation
$ E : \cm \rightarrow \cm $ such that the inclusion has
 depth 2, we show quite explicitely  how $\cn $ can be viewed
as the fixed
point algebra of $\cm$ w.r.t. an outer action of a compact Kac-algebra
acting on $\cm$.
This gives an alternative proof, under this special setting, of a more
general result  of
M. Enock and R. Nest, [E-N],  see also S. Yamagami, [Ya2].
\end{abstract}

\section{Introduction}

In the algebraic approach  to quantum field theory as proposed by R. Haag
et al. , [Haa], one has the natural inclusion of the observable algebra
lying inside the field algebra.
Due to the beautiful results of S. Doplicher and J. Roberts, [Do-Ro], at least
in
four spacetime dimension this inclusion can be described by the outer action of
a compact group on the field algebra such that the observables are
 the fixed point elements under this
 action.
In lower dimension the question how in general the observable algebra sits
inside the field algebra is still open ( symmetry problem in quantum field
theory ).

The basic inclusion one starts with is of the type
$$
\rho(\ca)\subset \ca,
$$
where $\rho$ is a localized endomorphism and $\ca $ the quasi local
$C^*$-algebra
of observables. Here $\rho$ represents a charged particle of the theory.
In this framework antiparticles are also described by localized
endomorphisms, the so called conjugate endomorphisms $\bar{\rho}.$
They are characterized by the existence of an observable intertwiner
$R_{\bar{\rho},\rho} \in \ca,$ such that
$$
 R_{\bar{\rho},\rho}  \in (id,\bar{\rho}\circ \rho).
$$
Due to the result of R. Longo, [Lo2], the existence of such an intertwiner is
equivalent to the existence of a conditional expectation
$E:\ca \rightarrow \rho(\ca),$ see also section 2, where we review some of
his results.

In order to get a better understanding of how things may look
like we investigate a simplified situation, i.e. we consider
 an irreducible inclusion of
infinite von-Neumann-factors $\cn \subset \cm$ of depth 2, see the definition
below,
together with a faithful conditional expectation $E : \cm \rightarrow \cn$.
It has first been claimed by A. Ocneanu, [Oc], and is meanwhile well known
 , [Lo1,Sz,Da], that at least in the finite index case
the  inclusion stems from the outer action of a Hopf-algebra on $\cm$.
In the quantum field theoretic setting of S. Doplicher and J. Roberts,
this would imply that the concept of group symmetry gets replaced by a
``quantum group symmetry'', where $\cm$ would be the field algebra and
$\cn$ the observable algebra.

It turns out that Longo's  intertwiner is extremely useful
in order to identify the Hopf algebra structure together with its dual,
the convolution product, Fouriertransform and Haar state etc. and the
formulae of  A. Ocneanu, [Oc1,2], are easiliy rephrased in terms which
also work in the infinite index case.

The plan of our paper is as follows. Given $E:\cm \rightarrow \cn$ and
Longo's canonical endomorphism $\gamma:\cm \rightarrow \cn$ we review
in section 2 some basic results of R. Longo, [Lo2,3] , on the Jones
triple $\gamma(\cm)\subset \cn \subset \cm$ and the associated
 intertwiner space $(id, \gamma).$
We also introduce the relative commutants $\ca_{i,j}=\cm_{i-1}' \cap
\cm_j,$ where $\cm_{2n}=\gamma^{-n}(\cm)$ and
$\cm_{2n-1}=\gamma^{-n}(\cn),$
and we discuss Ocneanu's depth 2 condition, [Oc1,2], in terms of a
Pimsner-Popa basis $\li \in \cao.$
In section 3 we use these data to define a Fourier transform $\cf:\cao
\rightarrow \cae$ and a convolution product * on $\cao.$
In section 4 we construct the coproduct on $\cao,$ the Haar weights on
$\cao$ and $\cae$ and prove that $\cf$ extends to a unitary on the
associated $L^2$-spaces.
We also have natural candidates for involutive antipodes on $\cao$
and $\cae,$ respectively, anticipating that $\cao$ and $\cae$ become a
pair of dual Kac algebras.
In section 5 we construct a natural Hopf-module right action of
$\ca_{0,1}\equiv \gamma^{-1}(\cae)$ on $\cm_0$ such that $\cm_1=\cm_0 \lhd
\ca_{0,1} $ becomes a crossed product.
In section 6 we construct a multiplicative unitary and use the results
of S. Baaj and G. Skandalis, [B-S], to complete the proof of our setting.

As a note on our terminology we remark that we talk of Kac algebras,
since the antipodes are involutive. We also call $\ca_{0,1}$ a compact Kac
algebra, since it contains the Jones projection $e_1$ as a two-sided
integral and therefore generalizes the role of a compact symmetry
group ( or rather its convolution algebra ). This agrees with [Ya3],
but opposes [Po-Wo, E-S], where $\ca_{0,1}$ would be called ``discrete''.

 During our investigation we got
a preprint of M. Enock and R. Nest, [E-N], where a much  more general
analysis is given.
They look at arbitrary irreducible depth 2 inclusions of factors
fulfilling a certain regularity condition and also construct a
multiplicative unitary.

In view of their work we do not claim to get any exiting new result.
Nevertheless, since their formulas are often quite involved and
implicit, it
 might be helpful to have an alternative approach  to our
 special case. Another advantage of our methods
is that many of the formulae we derive work quite well also for reducible
inclusions of depth 2. But this is still under investigation, [N-W] .
We hope to come back to this in the near future.

\section{Preliminaries}

We start with introducing the notations and, for the readers convenience,
with reviewing some of the results obtained by R. Longo, [Lo2,3], which
we will use in the following.

\noindent Let $\cn \subset \cm $ be an irreducible inclusion of von-Neumann
factors
with separable predual, acting on a Hilbert space $\tilde{\ch}$.
Let $\om $ be a common cyclic and separating vector .
Denote $\jm , \jn$ the associated modular conjugations and
$$ \gamma := Ad \ \jn \jm $$
the canonical endomorphism of Longo, see [Lo4].
We consider the Jones tower
$$ \cn = \cm_{-1} \subset \cm = \cm_0 \subset \cm_1 \subset \cm_2 \ . \ . $$
where
$$ \cm_{2n-1} := \gamma^{-n} ( \cn ) \ \ \ \ , \ \
\cm_{2n} = \gamma^{-n} ( \cm ) ,  $$
see [Lo2,3].
We will assume the existence of a faithful conditional expectation
$$ E_0 : \cm \rightarrow \cn. $$
By the work of R. Longo, [Lo2], this assumption is equivalent to the
existence of an intertwiner $w_{-1} \in (id, \gamma), \ \ w_{-1} \in \cn:$
Take the state
$$ m \mapsto < \om, E_1 (m) \om >, \ \ \ m \in \cm $$
on $\cm.$ By modular theory, [St-Zs], we can represent this state uniquely
by a vector $ \xi \in P^{\natural} ( \cm ,\om),$ where
$ P^{\natural} ( \cm ,\om) $
is the natural positivity cone of $(\cm , \om).$
Then the orthogonal projection
$$ e_1 : \ch \rightarrow \overline{\cn \xi} $$
is called the Jones projection associated with the conditional expectation
$E_0.$
{}From the definition  we get
$$ [ e_1 , \jm ] = 0. $$
Let
\bea
w_{-1}' : \cn \om & \rightarrow & \cn \xi \nn \\
      n \om & \mapsto & n \xi \ \ \ n \in \cn .
\ea
Then $w_{-1}' $  obviously defines an isometry in $\cn ' $ obeying
$$ w_{-1}'^* w_{-1}' = {\bf 1} , \ \ w_{-1}' w_{-1}'^* = e_1 . $$
Moreover, as was noticed by R. Longo, [Lo2], we have the relation
 $$ \jm w_{-1}' \jn = w_{-1}' . $$
As a Corollary we get, see [Lo2],

\begin{corollary}
Let $ w_{-1} := Ad \jn (w_{-1}') \in \cn.$
Then $w_{-1}$ is an isometry obeying
$$
w_{-1}^* w_{-1}={\bf 1}, \ \ \ w_{-1}w_{-1}^*=e_{-1},
$$
$$
w_{-1} n = \gamma(n) w_{-1} \ \ \forall n \in \cn.
$$
\end{corollary}

\proof
The first line follows immediately from the definitions. To prove
the intertwiner property let $n \in \cn.$ Then
\bea w_{-1} n & = & \jn v' \jn n = \jn (\jm v' \jn ) \jn n \nn \\
              & = & \jn \jm v' n = \jn \jm n \jm \jn \jn \jm v' \nn \\
              & = & \gamma ( n) \jn v' \jn = \gamma ( n) w_{-1} \nn
\ea
\hfill $\Box$

As an important tool we need the notion of a Pimsner-Popa basis of $ \cm $
over $\cn$ via $E_0,$ see [Pi-Po]. Let $\{ m_j \}_j \in \cm$ be a family
of elements in $\cm$ satisfying the conditions
$$
1) \ \ \{ m_j e_1 m_j^* \}_j \ \mbox{ are mutually orthogonal projections}
$$
$$
2) \ \ \sum_j m_j e_1 m_j^* =1 .
$$
Then the family $\{ m_j \}_j \in \cm$ is called a Pimsner-Popa basis associated
with $E_0 :\cm \rightarrow \cn.$
Using a Gram-Schmidt orthogonalization procedure one proves the existence
of such a basis.

We need a special case of a Pimsner-Popa basis.
Denote $$ \mu_0 := E_0 |\cao \rightarrow {\bf C} .$$
A state on $\cao $ can be viewed as a very special case of a conditional
expectation, where
the subalgebra of $ \cao$  is ${\bf C 1} \subset \cao.$
Therefore we may first construct a Pimsner-Popa basis associated with
$(\cao, \mu_0).$

Let $\ch $ be the GNS-Hilbert space associated with $(\cao, \mu_0).$
 Using a Gram-Schmidt orthogonalization procedure we
 choose elements
$$ \li \in \cao \subset \cm $$
with
$$ \mu_0 ( \li^* \lj ) = \delta_{ij}, $$
$ \lam_0 = {\bf 1} ,$ such that they yield an orthonormal basis in $\ch.$
For this special situation such a  basis is a Pimsner-Popa basis,
see [Pi-Po, He-Oc].

Now let the inclusion $ \cn \subset \cm $  have depth 2, which says
that $\cn' \cap \cm_2 $ is a factor or
$$ {\bf C} = \cn ' \cap \cm \subset \cn' \cap \cm_1 \subset \cn ' \cap \cm_2
$$
is again a Jones tower, see [He-Oc, E-N].
Therefore we get
\be
\sum_i \li e_1 \li^* = {\bf 1},
\ee
where the limit is taken in the strong topology on $\tilde{\ch}.$
 Hence the above basis also gives  a Pimsner-Popa
basis for $\cm ,$
$$
\mbox{ depth 2} \Rightarrow \li \in \cao \subset \cm \mbox{ is a Pimsner-Popa
basis for }
 E_0 : \cm \rightarrow \cn. $$

Next we define shifted conditional expectations,
Jones projections and intertwiners
$$ E_{2n} := \gamma^{-n} \circ E \circ \gamma^{-n} :
\cm_{2n} \rightarrow \cm_{2n-1}                   $$
$$
e_{2n+1}:= \gamma^{-n}(e_1) \in \cm_{2n+1}, \
$$
$$
w_{2n-1} :=\gamma^{-n} (w_{-1}) \in \cm_{2n+1} .
$$
Then
$$ w_{2n+1} x = \gamma (x) w_{2n+1}, \ \  \forall x \in \cm_{2n+1}. $$

Our basic objects of interest are the relative commutants
 \be
 \ca_{i,j } := \cm_{i-1} '\cap \cm_j .
\ee

In particular we have
$$ \ca_{-1,0}:= \cm \cap \gamma(\cm')=\cm \cap Ad \ \jn (\cm)$$
and
$$ \ca_{0,1} := \cm_1 \cap \cn' = \mbox{ Ad }\jm \ (\cn') \cap \cn'.$$
We will show below that in the irreducible depth 2 case $\cao$
carries a natural $^*$-Hopf-algebra structure with dual algebra $ \cae.$
This can be generalized to all steps in the tower.
The antipodes on $\cao$ and $\cae$ will turn out to be involutive, hence
we are actually dealing with Kac algebras.

\noindent To identify the Haar state on $\cao$ we consider the restricted
conditional expectations
 $$E_{2n}| \ca_{2n-1,2n} \rightarrow \ca_{2n-1,2n-1}.$$
Since we will assume irreducibility, i.e.
$$ \cm \cap \cn'= {\bf C}.$$
we  have
$$\gamma^{-n} (\cm \cap \cn')= \ca_{2n,2n} = {\bf C}$$
and similarly $\ca_{2n-1,2n-1}={\bf C},$
and the restricted conditional expectations reduce to states.
We will identify $E_0 |\cao$ to be the Haar state of $\cao$, when $\cao$
is equipped with the Hopf algebra structure mentioned above.

\vspace{.3cm}
We close this section with a brief review of some useful identities, which
hold for irreducible inclusions ( not necessarily of depth 2) and are all due
to R. Longo, [Lo2].

\begin{lemma} [Lo2]
$E_0 : \cm \rightarrow \cn $ is given by
$m \mapsto w_{-1}^*\gamma(m) w_{-1} .$
\end{lemma}

\proof
One easily checks that $m \mapsto w_{-1}^*\gamma(m) w_{-1} $ defines
a conditional expectation from $ \cm \rightarrow \cn.$
By irreducibility the conditional expectation is unique.
\hfill $\Box$

\vspace{.5cm}

This later result leads to

\begin{corollary}
$${\bf 1}\mu_0 (a) = w_{-1}^* \gamma(a) w_{-1} = w_{1}^* a w_{1}
= w_{-1}^* a w_{-1}
\ \ \ \forall a \in \cao.$$
\end{corollary}
\proof
 For $a \in \cao$ we have
$E_0(a)=\gamma^{-1} \circ E_0(a) \in {\bf C}$ and therefore
$$
{\bf 1}\mu_0 (a) =
 w_{-1}^* \gamma(a) w_{-1}
 =  \gamma^{-1} ( w_{-1}^* \gamma(a) w_{-1} ) = w_{1}^* a w_{1} .
$$
To show that  $$ w_{-1}^* a w_{-1} \in \cm \cap \cn' = {\bf C}. $$
let $m_{-1} \in \cn.$ Then
$$  w_{-1}^* a w_{-1} m_{-1} =  w_{-1}^* a \gamma (m_{-1}) w_{-1}.$$
But $ \gamma(m_{-1}) \in \gamma(\cn)$ commutes with
$\  a \in \cm \cap \gamma(\cn'),$ and
we get
$$ \hspace{3cm} = m_{-1}  w_{-1}^* a w_{-1}. $$

Using the intertwining property $ w_{-1}w_1 = w_1 w_1$  we finally get
\bea
w_{-1}^* a w_{-1} & = & w_{1}^* w_{-1}^* a w_{-1}  w_{1}  =
w_{1}^* w_{1}^* a w_{1}  w_{1} \nn \\
& = & w_{1}^* a w_{1} w_{1}^*  w_{1}   = w_{1}^* a w_{1} .\nn
\ea
\hfill $\Box$

We will frequently make use of the different descriptions of the state
$\mu_0.$ Sometimes we will implicitely identify
${\bf C 1} \subset \cao \cong {\bf C}$ and simply write
$\mu_0(a)=w_{-1}^*aw_{-1}$ etc..

The last corollary we need is
\begin{corollary}
For $ a \in \cao $ we have
$ w_{1}^* w_{-1}^* a w_{1} = \mu_0 (a) w_{1}^* . $
\end{corollary}

\proof
 For $ a \in \cao $ we have
$$
w^*_1w^*_{-1} a w_1 = w_1^* w_1^* a w_1= w_1^* \mu_0(a). $$
\hfill $\Box$

\section{The Fourier transform and the convolution product}

To consider a special case let us  assume we already knew that  $ {\cal
L}(\ch)\supset \cm = \cn \lhd G $
is a crossed product, where $G $ is
an abelian compact group
acting outerly on $\cn.$
In this case the dual group $\hat{G}$ would naturally act on $\cm$ and we would
have a unique faithful conditional expectation from
$\cm$ to $\cn$ given by the projection onto the $\hat{G}$-invariant part,
i.e. by avaraging over $\hat{G}$ w.r.t. the
Haar measure on $\hat{G}$. $G$ acts on
$\cn$ and we can decompose $\cn$ into a direct sum of irreducible
representations of $G$,
$$
\cn = \oplus_{\hat{g}\in\hat{G}} \cn_{\hat{g}} .
$$
where for simplicity we have also assumed $\hat{G}$ to be finite.
Due to the assumption of $\cn$ being an infinite factor, we have a
spatial isomorphism
$$
\cn \cong \cn_{\hat{g}}.
$$
Denote $u_{\hat{g}}$ the associated intertwiners. Again, according
to the assumption of $\cn$ being an infinite factor, we have
$$\cn \cong \cn^{\sharp \hat{G}}. $$
It is not hard to see that
$$w_{-1} \cong \oplus_{\hat{g}\in\hat{G}} u_{\hat{g}} .$$

Moreover, by Landstad's result, see [Pe], we would have
$$ (\cn \lhd G \lhd \hat{G} ) \cap \cn' = L^{\infty} (G, \mu ), \ \ \
 (\cn \lhd G \lhd \hat{G} \lhd G ) \cap \cm' = L^{1} (G, \mu )
$$
where $L^{\infty} (G, \mu )$ is the Fourier algebra, $L^{1} (G, \mu ) $
the convolution algebra over $G,$
$\mu$ the Haar state.

In terms of our Jones tower we would have
$$
 \cn \lhd G \lhd \hat{G}   \cong \cm_1, \ \ \
 \cn \lhd G \lhd \hat{G} \lhd G \cong \cm_2.
$$
Using the isomorphism $\gamma$ we can reformulate the above observation as
$$\cae = \gamma(\cn' \cap \cm_1) \cong  L^{\infty} (G, \mu ), \
$$
$$
  \cao = \gamma ( \cm' \cap \cm_2 ) \cong L^1 (G, \mu ).
$$
Moreover, by Takesaki duality, see [St], we would have
$$
\cm_2 \cong \cm \otimes {\cal L}(L^2(G,\mu))
$$
and under this identification the inclusion $\cn \subset \cm_2$
would become
$$
\cn \otimes {\bf 1}_{L^2(G,\mu)} \subset
\cm \otimes {\cal L}(L^2(G,\mu)).
$$
Hence $\ca_{0,2} \equiv \cn' \cap \cm_2 =  {\cal L}(L^2(G,\mu))
$
which would imply Ocneanu's depth 2 condition.
It was A. Ocneanu, see [Oc1,2], who first noticed that this picture
generalizes to the case where $G$ and $\hat{G}$ are replaced by an
arbitrary dual pair of finite dimensional Kac algebras.
He also claimed that all irreducible finite index and
depth 2 inclusions of factors arise in this way.
Proofs of this statement have later been provided by [Lo, Szy, Da].
A rather extensive generalization to infinite index inclusions has recently
been given by M. Enock and R. Nest, [E-N].
The aim in our paper is to give an alternative identification of $\cao$ and
$\cae$ as a pair of dual Kac algebras, such that $\ca_{i,i+1}$  naturally
acts on $\cm_i$ and
$$
\cm_{i+1} = \cm_i \lhd \ca_{i,i+1}.
$$

Let us start with some generalities.
To simplify the computation we will use as a concrete realization of the
GNS-Hilbert
space $\ch$ the subspace
\be
\ch \cong \overline{\cao w_{-1} \om } \subset \tilde{\ch} .
\ee
We will use Dirac's ket-vector notation to denote
\be
|a>:= aw_{-1} \om, \ \ a \in \cao
\ee
such that the left action of $\cao$ on $\ch$  reads
$$
 a|b> = |ab>, \ \ \ a,b \in \cao .
$$
By Corollary 3 we could also use $w_{1}\om$ instead of $w_{-1} \om.$
There is a simple relation between both realizations:

\begin{lemma}

\begin{enumerate}
\item $Ad \ \jn ( \cao) = \cao,$
\item  $\jn w_1 \om = w_{-1}\om.$
\end{enumerate}
\end{lemma}

\proof
 1) follows by the definition, $ \cao = \cm \cap Ad \ \jn (\cm ).$

\noindent For 2) notice that by definition we have
$$ w_1\om = \xi \in P^{\natural} ( \cm ,\om) $$
and therefore
\be
\jn w_1 \om = \jn \jm w_1 \om = \gamma ( w_1) \om = w_{-1} \om
\ee
\eot

Let us remark here that with
$\li $ also $\bar{\li} := \jn \li \jn \in \cao$ is a Pimsner-Popa basis:
\bea
\mu_0(\bar{\lambda_j}^* \bar{\lambda_i})& = &
<w_1 \om ,\bar{\lambda_j}^* \bar{\lambda_i} w_1 \om> \nn \\
& = & <w_1 \om , \jn \lambda_j^* \lambda_i \jn w_1 \om > =
< w_{-1} \om , \lambda_i^* \lambda_j w_{-1}\om > \nn \\
& = & \mu_0 ( \lambda_i^* \lambda_j ) = \delta_{i,j} . \nn
\ea
Especially we get
\bea
{\bf 1} & = & \sum_i \bar{\lambda_i}e_1 \bar{\lambda_i}^* \nn \\
& = & \jn ( \sum_i \li \jn e_1 \jn \li^* )\jn . \nn
\ea
Now we use $e_{-1} = \gamma (e_1)$ and $\mbox{Ad } \jm (e_1) =e_1,$
to get $ e_{-1} = \jn e_1 \jn,$ which yields
\be
 \sum_i \li  e_{-1}  \li^* = {\bf 1}.
\ee

We also immediately get the

\begin{corollary}
Assume depth 2. Then $\ca_{-2,0} = \cm \cap \gamma(\cn)' $ can
naturally be identified with
${\cal L} ( \ch)$ such that $\cao \subset \ca_{-2,0} $ acts by the
GNS-representation on $\ch.$
\end{corollary}

\proof
 By the assumption of depth 2 we get
$$
{\bf C} = \ca_{-1,-1} \subset \cao \subset \ca_{-1,1}.$$
is a Jones tower. This implies that
$$
\{ ae_1 b | \ a,b \in \cao \} \subset \ca_{-1,1}
$$
is a dense *-subalgebra, see [Pi-Po, He-Oc, E-N].
Now we use again $e_{-1}=\jn e_1 \jn$ which yields
\be
\{ ae_{-1} b | \ a,b \in \cao \} \subset \mbox{Ad } \jn \ (\ca_{-1,1})
= \ca_{-2,0}
\ee
is a dense *-subalgebra.
Define
$$
\Lambda( ae_{-1} b^* ) := |a><b| \ \in {\cal L}(\ch),
$$
where we used the bra-ket notion for vectors and forms.
 $\Lambda $ can be continued to a normal morphism on $\ca_{-2,0}.$
For $a\in \cao \subset \ca_{-2,0}$ we rewrite, using the
Pimsner-Popa-basis $\li ,$
$$
a=\sum_i a\li e_{-1} \li^* $$
which yields
$$
\Lambda(a) =  \sum_i |a\li > <\li | = a \sum_i |\li><\li| = a
$$
where we used that $|\li >$ is an orthonormal basis of $\ch.$
This shows compatibility with the GNS-representation.
\eot

\noindent By definition we have
$$ \cae = \cn \cap \gamma(\cn ') \subset \ca_{-2,0} = {\cal L}(\ch) $$
and we identify $\cao$ resp. $ \cae $ with their images in ${\cal L}(\ch).$

As a first step towards identifying $\cao$ and $\cae$ as a dual pair
of Kac algebras we construct what will become the Fourier transform:
\bea
\cf : \cao & \rightarrow & \cae\nn \\
 a & \mapsto & \gamma (w_{-1}^*aw_1 )=w_{-3}^* \gamma (a) w_{-1}.
\ea
To see that $\cf (a) \in \cae$ first note that
obviously $ \cf(a) \in \cn.$ Let $n \in \cn.$ Then
\bea
\gamma(w_{-1}^* a w_1) \gamma ( n)  & = & \gamma(  w_{-1}^* a w_{1} n) \nn \\
& = & \gamma(w_{-1}^* a \gamma (n ) w_{1} ) \ \ \ \ \mbox{ ( by using the
intertwiner property of $w_1$)} \nn \\
& = & \gamma (w_{-1}^*\gamma(n) a w_1 ) \ \
\ \ (\mbox{since }a \in \gamma(\cn)' ) \nn \\
& = & \gamma (n w_{-1}^*aw_1 )  .
\ea
Therefore
$ \cf(a) = \gamma(n) \gamma( w_{-1}^* a w_1) \in \cn \cap \gamma(\cn) '
= \cae.$

Moreover $\cf$ is injective, since
$$ \gamma(w_{-1}^* a w_1) =0 \Rightarrow w_{-1}w_{-1}^* a w_1 w_1^*
= e_{-1} a e_1 =0
\Rightarrow e_{-1}a=0, $$
where the last implication is standard from Jones theory, since
$e_{-1}a \in \cm.$
Using $e_{-1} = \jn e_1 \jn , $ see the proof of Corollary 6,
 and $ \jn a \jn \in \cm $
by the first identity in Lemma 5, the same argument also gives $a=0,$
i.e. $\cf $ is injective. We will show below , see
Theorem 17, that $\cf$ has  dense range.

\begin{definition}
The injective linear map $ \cf :\cao  \rightarrow  \cae $
is called the Fourier transform.
\end{definition}

We will show below that this map is really the Fourier transform
of the underlying Kac-structure.

In the finite index case the notion of a Fourier transform
was introduced by A. Ocneanu, see [Oc1,2].
A simple computation shows that in the finite index case
his formula agrees
 with ours.
It is at this point where the language of infinite algebras
has the advantage of allowing a unified description which also works in the
infinite index case.

We also have  a natural convolution product * on $\cao:$
\bea
* : \cao \times \cao & \rightarrow & \cao \nn \\
        ( a,b) & \mapsto & a* b := w_{-1}^*a\gamma(b) w_{-1}
\ea
To see that indeed $a*b \in \cao, \ a,b \in \cao,$
let $m_{-2}\in \cm_{-2}\equiv \gamma(M).$  Then
\bea
w_{-1}^*a\gamma(b) w_{-1} m_{-2}
& = & w_{-1}^*a\gamma(b) \gamma(m_{-2}) w_{-1}
\ \ \ \mbox{( by the intertwiner property)} \nn \\
& = & w_{-1}^*\gamma(m_{-2}) a\gamma(b) w_{-1} \ \
(\mbox{since } a\gamma (b) \in \gamma (\cm_{-2})' )\nn \\
& = & m_{-2} w_{-1}^*a\gamma(b) w_{-1}. \nn
\ea
This proves $ a*b \in \cm \cap \gamma(\cm') = \cao.$

\begin{definition}
The product $*: \cao \times \ \cao \mapsto \cao$ is called the convolution
product on $\cao.$
\end{definition}

In order to have consistent notations we check

\begin{lemma}
$$
\cf ( a*b) = \cf (a ) \cdot \cf (b) \ \ \ \forall a,b \in \cao.
$$
\end{lemma}

\proof
 Using successively the intertwiner properties
of the $w_i'$s we compute
\bea
\cf (a*b) & = & \gamma (w_{-1}^* w_{-1}^* a \gamma (b) w_{-1} w_{1}) \nn \\
          & = & \gamma (w_{-1}^* w_{-3}^* a \gamma (b) w_{1} w_{1}) \ \
 \nn \\
          & = & \gamma (w_{-1}^* a w_{-3}^* w_1 b w_1 )
 \ \ (\mbox{ since } a\in \cm_{-2}' \subset \cm_{-3}')\nn \\
          & = &\gamma ( w_{-1}^* a w_1 w_{-1}^* b w_1 )
= \cf (a) \cdot \cf (b) \nn
\ea
\eot

The injectivity of the Fourier transform implies

\begin{corollary}
The convolution product is associative.
\end{corollary}

Note that the formulae for the Fourier transform and convolution product
also make  sense if the underlying inclusion does not have depth 2, as was
first pointed out by A. Ocneanu, [Oc1,2]. Moreover, even if the inclusion
$\cn \subset \cm$ is not irreducible but of finite index there exists
 a unique minimal conditional expectation $E_0:\cm \rightarrow \cn,$ see
 [Ko,Lo]. Taking as above the uniquely associated intertwiner $w_{-1}\in \cn$
 one gets natural generalizations of the above notions to the reducible cases.
 A more detailed analysis of these generalizations is still under
 investigation, [N-W].

\section{Coproducts, Haar weights and the Plancherel Formula}

In this section we start with the definition of the coproduct
\bea
 \Delta : D(\Delta )\subset \cao & \rightarrow & \cao \otimes \cao \nn \\
                            a & \mapsto & a^{(1)} \otimes a^{(2)} \nn
\ea
where we used the Sweedler notation implying a summation convention in
$\cao \otimes \cao.$
The coproduct is defined as the $L^2(\cao, \mu_0)$-transpose of the
convolution product, i.e.
\be
\mu_0 (a(b*c))=:(\mu_0 \otimes \mu_0)((a^{(1)} \otimes a^{(2)})(b\otimes c)),
\ \ \ \forall b,c \in \cao .
\ee
and
$
 a \in D(\Delta)  \mbox{ iff } $
$
\exists a^{(1)} \otimes a^{(2)} \in \cao \otimes \cao $ such that
equation (12)
holds .
Here we have not assumed the depth 2 condition.
Faithfulness of $\mu_0$ implies that $\Delta$ is well defined and
coassociative, since the convolution product * is associative.
In the finite index case, where we don't have to worry about domains of
definitions, we get

\begin{lemma}
\bea
\Delta : \cao &\rightarrow & \cao \otimes \cao  \nn \\
a & \mapsto & \sum_{i,j} \lambda_i \otimes
\lambda_j w_{-1}^*\gamma(\lambda_j^*) \cf (\lambda_i)^* \gamma(a ) w_{-1}.
\ea
\end{lemma}

\proof
 Let $ a,b,c \in \cao .$ Then
\bea
\mu_0 ( b*c,a) & = & w_{-1}^*(w_{-1}^*b^* \gamma(c^*)w_{-1})aw_{-1} \nn \\
& = & w_{-1}^* w_{-3}^*b^* \gamma(c^*)w_{-1}aw_{-1} \nn \\
& = & w_{-1}^* b^* w_{-3}^* \gamma(c^*)w_{-1}aw_{-1} . \nn
\ea
Now we use the equation (2)  and Corollary 1 to conclude
\bea
\hspace{1cm} & = &
\sum_i w_{-1}^*b^*(\lambda_iw_{-1} w_{-1}^*\lambda_i^*)
w_{-3}^* \gamma(c^*)w_{-1})aw_{-1} \nn \\
& = &
\sum_i (w_{-1}^* b^* \lambda_i w_{-1}) w_{-1}^*
w_{-3}^*  \lambda_i^* \gamma(c^*) w_{-1} a w_{-1} \nn \\
& = &
\sum_i \mu_0(b^* \lambda_i) \mu_0((\lambda_i^* * c^* ) a ). \nn
\ea
We observe
\bea
\mu_0((\lambda_i^* * c^* ) a ) & = &
w_{1}^*
w_{-1}^*  \lambda_i^* \gamma(c^*) w_{-1} a w_{1}  \nn \\
& = &
w_{1}^* c^*
w_{1}^*  \lambda_i^*  w_{-1} a w_{1}  . \nn
\ea
Now we use again that $ \lambda_j$ is a Pimsner-Popa basis for
$E_0:\cm \rightarrow \cn$,
\be
{\bf 1} = \sum_j \lambda_j e_1 \lambda_j^* =
\sum_j \lambda_j w_1 w_1^*  \lambda_j^* =
\sum_j \lambda_j w_{-1}^* \gamma(\lambda_j^* ) w_1
\ee
and plugging this into the above formula yields
\bea
\mu_0((\lambda_i^* * c^* ) a ) & = &
\sum_j w_{1}^* c^*  \lambda_j w_{-1}^* \gamma(\lambda_j^* ) w_1
w_{1}^*  \lambda_i^*  w_{-1} a w_{1}  \nn \\
& = &
\sum_j w_{1}^* c^*  ( \lambda_j w_{-1}^* \gamma(\lambda_j^* )
w_{-1}^*  \gamma(\lambda_i^*)  w_{-3} \gamma(a) w_{-1} w_1  \nn \\
& = &
\mu_0(c^* ( \sum_j \lambda_j w_{-1}^* \gamma(\lambda_j^* )
\cf (\lambda_i)^* \gamma(a) ) w_{-1})) \nn
\ea
\eot

Under the assumption of depth 2
it will follow from section 6 that the formula
for the coproduct is also valid in the infinite index case with
$D(\Delta)=\cao$ and moreover that it is an algebra morphism, i.e.
$\Delta(ab)=\Delta(a)\Delta(b), a,b \in \cao.$
Here we only note that ${\bf 1} \in D(\Delta) $ and
$\Delta({\bf 1})= {\bf 1} \otimes {\bf 1}$ and $ \Delta(a)^* = \Delta (a^*),
a \in D(\Delta),$ which follows from
\begin{lemma}
$$ i) \ \ \ \mu_0(a*b)=\mu_0(a)\mu_0(b), \ \ a,b \in \cao  $$
$$ii) \ \ \ (a*b)^*=a^**b^*, \ \ a,b \in \cao $$
\end{lemma}

\proof
\noindent
i)  Using Corollary 3 together with
$\mu_0(b)=\gamma(\mu_0(b))\in {\bf C1}, b \in \cao,$ we have
for $a,b \in \cao$
\bea
\mu_0(a*b) & = & w_{-1}^* w_{-1}^* a \gamma(b) w_{-1} w_{-1}
= w_{-1}^*w_{-3}^* a \gamma(b) w_{-3}w_{-1} \nn \\
& = & w_{-1}^* a \gamma(w_{-1}^*bw_{-1})w_{-1} =
w_{-1}^* a \gamma(\mu_0(b))w_{-1} \nn \\
& = & \mu_0(a)\mu_0(b). \nn
\ea

ii) Follows immediately from the definition of the convolution product, since
 $a\gamma(b)=\gamma(b)a, a,b \in \cao.$
\eot

Clearly i) implies ${\bf 1} \in D(\Delta) $ and
$\Delta({\bf 1})= {\bf 1} \otimes {\bf 1},$ whereas
$ \Delta(a)^* = \Delta (a^*), a \in D(\Delta),$ follows from ii),
provided $\mu_0 $ is a trace, which has been shown by S. Yamagami, [Ya1],
see also [Ya2]. In the remark to Proposition 19 below we state a
conjecture, which
also would imply the trace property of $\mu_0$ and which in the
finite index case is well known to hold.

Next we show that $\mu_0 $ is left and right invariant w.r.t. $\Delta$
and may hence be called a Haar state.
\begin{lemma}
$$(\mbox{id } \otimes \mu_0) (\Delta(a))=
  (\mu_0 \otimes \mbox{id } )(\Delta(a))= \mu_0(a) {\bf 1} ,
\ \ a\in D(\Delta).$$
\end{lemma}

\proof
\noindent From the definition of the convolution product and
Corollary 3 we have for all $ b \in \cao $
$$ b*{\bf 1} = {\bf 1}*b = \mu_0(b) {\bf 1}. $$
Hence, we have for $a\in D(\Delta), b\in \cao$
$$
(\mu_0 \otimes \mu_0)(({\bf 1} \otimes b) \Delta(a))
= \mu_0(({\bf 1}*b)a) = \mu_0(b) \mu_0(a),
$$
$$
(\mu_0 \otimes \mu_0)((b \otimes {\bf 1} ) \Delta(a))
= \mu_0((b*{\bf 1})a) = \mu_0(b) \mu_0(a),
$$
from which the Lemma follows by faithfulness of $\mu_0.$
\eot

We postpone the construction of the antipode on $\cao$ to a later stage,
see  Definition 18 below.

Next we remark that by Lemma 9 $\cae$ naturally appears as the dual
algebra of $(\cao,\Delta),$ where the pairing is given by
$$
<a,\cf(b)>:= \mu_0(ab), \ \ \ a,b \in \cao.
$$
In particular, since $\cf({\bf 1})=e_{-1},$ we get
$$
\mu_0(a)=<a,e_{-1}>,
$$
and hence $e_{-1}$ is the two-sided integral in $\cae.$
Indeed, we also have a counit $\hat{\varepsilon}$ on $\cae$ given
on $\hat{a}\in \cae$ by
$$
\hat{\varepsilon}(\hat{a})=w_{-1}^* \hat{a} w_{-1} \in \ca_{-1,-1}= {\bf C}.
$$
Using equation (9) and Corollary 3 one easily checks that
$$
\hat{\varepsilon}(\cf(b))=<{\bf 1}, \cf(b)>=\mu_0(b),
$$
and Lemma 13 implies for all $\hat{a}\in \cae$
$$
\hat{a} e_{-1}=e_{-1} \hat{a} = \hat{\varepsilon}(\hat{a})e_{-1}.
$$
\vspace{1cm}

We now construct the Haar weight on the dual algebra
$\cae \cong \ca_{0,1}.$
Let us assume finite index for the moment.
Then we can treat the inclusion $ \cm \subset \cm_1$  on the same footing as
$\cn \subset \cm$ and define an intertwiner $w_0 \in \cm$ ( and hence
$w_{2n}\in \cm_{2n}$ ) analogously as $w_{-1}$ ( and hence $w_{2n-1}$ ).
We just  have to start from the conditional expectation
$E_1 : \cm_1 \rightarrow \cm $ associated to $E_0 : \cm \rightarrow \cn$ via
Jones theory.
In particular we get
\be
 w_0 m = \gamma(m) w_0, \ \  \ m \in \cm
\ee
and the normalization of $w_0$ could be fixed by requiring
\be
w_{-1}^*w_0 =
w_1^* w_0={\bf 1} ,
\ee
 see [Lo1], where the above structure has been called an
`` irreducible Q-system''.
Inspired by Lemma 2 we then get a state $\mu_1 $ on $ \ca_{0,1} \cong
\cae $ by
$$
\mu_1 (a) = w_0^* \gamma(a) w_0, \ \ \ a \in  \ca_{0,1}
$$
which apart from the normalization  coincides with the
restriction $E_1| \ca_{0,1}.$

We now  show how these ideas carry over to the infinite index case.
To this end we will derive formulas for $w_0$ and $E_1$ for the finite
index case and show that the resulting formula for $\mu_1$ also
makes sense in the infinite index case.

Using the depth 2 condition we know that $\li \in \cao \subset \cm $
is also a Pimsner-Popa basis for $E_0:\cm \rightarrow \cn.$
Hence
$$
{\bf 1} = \sum_i \li e_1 \li^* \in \cm_1 .
$$
Using $e_1=w_1 w_1^*$ and $mw_1^*=w_1^* \gamma(m) , \ m \in \cm,$ we get
$$
\hspace{0.5cm} = w_1^*(\sum_i \gamma(\li)w_{-1}\li^*),
$$
an identity which we already used in the proof of Lemma 11.
Hence we are led to the  Ansatz
$$
w_0 = \sum_i \gamma(\li)w_{-1}\li^* ,
$$
which is obviously well defined in the finite index case.
With this Ansatz we compute
\bea
w_{-1}^*w_0  &=& \sum_i w_{-1}^* \gamma(\li)w_{-1}\li^* \nn \\
& = & \sum_i E_0(\li) \li^* = {\bf 1}, \nn
\ea
where we have used Lemma 2. This proves (16).

To check (15) let $m \in \cm.$
Then
\bea
w_0 m & = & \sum_i \gamma(\li)w_{-1} \li^* m \nn \\
&=& \sum_{i,j}  \gamma(\li)w_{-1} E_0(\lambda_i^* m  \lambda_j)
\lambda_j^* \nn \\
& = &
\sum_{i,j}  \gamma(\li)w_{-1}w_{-1}^* \gamma(\lambda_i^*  m  \lambda_j) w_{-1}
\lambda_j^* \nn \\
& = &
\sum_i  \gamma(\li e_1 \lambda_i^* )\gamma( m) \sum_j \gamma(\lambda_j )w_{-1}
\lambda_j^* \nn \\
& =& \gamma(m) w_0, \nn
\ea
where we have used that $\li$ is a Pimsner-Popa basis for $E_0$ in the second
line and Lemma 2 in the third line.
This proves (15).

Motivated by Lemma 2 and the above computation we now define
\begin{definition}
\bea
E_1 : \cm_1^+ & \rightarrow & \bar{\cm^+} \nn \\
a &\mapsto & w_0^*\gamma(a)w_0=
\sum_{i,j} \lambda_i w_{-1}^* \gamma(\lambda_i^* a
\lambda_j)   w_{-1}^*  \lambda_j^*
\ea
\end{definition}

Then we get in the finite index case:
\begin{lemma}
Assume the depth 2 condition. Then
$ E_1$ is a normal faithful conditional expectation, obeying
$$ E_1(me_1m^*) = mm^* \ \ \ m \in \cm$$
\end{lemma}

\proof
\noindent
Let $m=\sum_{l=1}^n \lambda_l n_l \in \cm, \ n_l \in \cn.$
Then
$$ me_1 m^* \in \cm_1^+ ,
$$
and we compute
\bea
E_1(me_1m^*) & = &
\sum_{i,j,l,l'} \lambda_i w_{-1}^* \gamma(\lambda_i^*
\lambda_l n_l e_1 n_{l'}^* \lambda_{l'}^*
\lambda_j)   w_{-1}^*  \lambda_j^*  \nn \\
& = &
\sum_{i,j,l,l'} \lambda_i  \gamma( w_1^*\lambda_i^*
\lambda_l n_l w_1 w_1^* n_{l'}^* \lambda_{l'}^*
\lambda_j w_1)     \lambda_j^*  . \nn
\ea
Using the intertwiner property, Corollary 3 and $ \lambda_i$
being a Pimsner-Popa basis we get
\bea
\hspace{2cm} & = &
\sum_{i,j,l,l'} \lambda_i  \gamma( w_1^*\lambda_i^*
\lambda_l w_1) n_l  n_{l'}^* \gamma( w_1^* \lambda_{l'}^*
\lambda_j w_1)     \lambda_j^* \nn \\
& = & \sum_{l,l'}
\lambda_l   n_l  n_{l'}^*      \lambda_{l'}^* = mm^* .
\ea
Now the span of $\{ me_1\hat{m} \in \cm_1 | m,\hat{m} \in \cm \} $ is $\cm_1$
and the proof is finished.
\eot

Note that irreducibility implies uniqueness of such a conditional
expectation.

{\bf Remark}
\noindent
It is easy to see that in the infinite index case  $E_1$
is also well defined as an operator valued weight.
Normality follows easily by the above formulas:

\noindent
Let
\bea
E_1^{I^{(N)}} : \cm_1^+ & \rightarrow & \bar{\cm^+} \nn \\
a &\mapsto &
\sum_{i,j \in I^{(N)}} \lambda_i w_{-1}^* \gamma(\lambda_i^* a
\lambda_j)   w_{-1}^*  \lambda_j^*  \nn
\ea
with $I^{(N)} \subset I$ a finite subset.
Then $E_1^{I^{(N)}}$ is obviously normal and therefore also
$$
E_1 := \overline{\lim}_{I^{(N)} \subset I} E_1^{I^{(N)}}.
$$
On finite linear
combinations $m=\sum_{l=1}^n \lambda_l n_l $ with $n_l \in \cn$ the proof of
Lemma 15 also applies to the above limit, showing
$ E_1(me_1m^*) = mm^* .$
Such finite linear combinations are dense in $\cm $ and we get
semifiniteness of $E_1.$
Conversely one could take the property
$$ E_1(me_1m^*) = mm^* \ \ \ m \in \cm$$
as the starting point of a definition, as was first done in [B-D-H].

\vspace{1cm}
In our analysis we don't need the specific  properties of the operator valued
weight,
so we will skip these points.
We now show how the definition of $\mu_{-1}=E_1\circ\gamma^{-1}|\cae$
generalizes to the infinite index
case.
To this end we use the following identities, valid in the finite index
case for all $x\in \cae,$ similarly as in Corollary 3:
\bea
{\mu_{-1}}:\cae^+ &\rightarrow& \bar{\bf R}^+ \nn \\
 x^* x & \mapsto & E_1(\gamma^{-1}(x^*x))
= w_0^*(x^*x)w_0 \nn \\
& & =  w_0^*( \gamma^{-1} ( x^*x))w_0 \nn \\
& & =
\sum_{i,j} \lambda_i w_{-1}^* \gamma(\lambda_i^* )
\gamma^{-1}(x^*x) \gamma(\lambda_j)w_{-1} \lambda_j^* .\nn
\ea
Now $\gamma^{-1}(x^*x) \in \ca_{0,1} \subset \cn'$ and
$ \gamma(\lambda_i) \in \gamma(\cm) \subset
\cn$ and we conclude
\be
 \mu_{-1}(x^*x) =
\sum_{i,j} \lambda_i w_{-1}^* \gamma(\lambda_i^* )
 \gamma(\lambda_j)w_{-1}\gamma^{-1}(x^*x) \lambda_j^*
= \sum_i \lambda_i \gamma^{-1}(x^*x) \lambda_i^*.
\ee
The last expression is the important one, since it
  also makes sense in the infinite index case.

\begin{definition}
On $\cae$ let $\mu_{-1}$ be the operator valued weight
\bea
{\mu_{-1}}:\cae^+ &\rightarrow& \bar{\cae}^+ \nn \\
 x^* x & \mapsto & = \sum_i \lambda_i \gamma^{-1}(x^*x) \lambda_i^* . \nn
\ea
\end{definition}

We now show that $\mu_{-1}$ is in fact a good Ansatz for the Haar weight on
$\cae,$ since
we have the following Plancherel formula:
\begin{theorem}
If the depth of $\cn \subset \cm $ is two, then
$\mu_{-1}$ defines a semifinite faithful normal weight and
the Fourier transform extends to a unitary
$$
\cf : L^2(\cao , \mu_0) \rightarrow L^2(\cae,\mu_{-1}),
$$
i.e. $\cf $ has dense range and
$$
\mu_{-1} ( \cf (a)^* \cf (a) ) = \mu_0 (a^*a)  \ \ \forall a \in \cao .
$$
\end{theorem}

\proof
\noindent Let $a\in \cao.$ Then
\bea
\mu_{-1} ( \cf (a)^* \cf (a) ) & = &
\sum_i \lambda_i w_1^* a^* w_{-1} w_{-1}^* a w_1 \lambda_i^* \nn \\
& = &
\sum_i \lambda_i w_1^* a^* e_{-1} a w_1 \lambda_i^* \nn \\
& = &
\sum_i  w_1^* \gamma (\lambda_i) a^* e_{-1} a \gamma (\lambda_i^*) w_1 \nn
\ea
by using the intertwiner property, and from
$ \gamma (\lambda_i) \in \gamma (\cm) \subset \cn, a\in \cn'$ we conclude
\bea
\hspace{2cm} &=&
\sum_i  w_1^* a^* \gamma (\lambda_i) e_{-1} \gamma (\lambda_i^*) a w_1 \nn \\
& = &
w_1^* a^* a w_1 = \mu_0 ( a^* a) \nn
\ea
where we used $ \lambda_i$ a Pimsner-Popa basis.
The Plancherel formula generalizes obviously to $ \mu_{-1} ( \cf ( a )^*
\cf (b) ) = \mu_0 ( a^* b), \ \forall a,b \in \cao.$

Next we show that
$
\{ \cf(a)^*\cf(b)|a,b \in \cao \} \subset \cae \ \ \mbox{ is dense. }
$
To this end we compute
\bea
\cf (a)^* \cf( b) & = &\gamma( w_1^* a^* w_{-1} w_{-1}^* b w_1 )
= \gamma(w_1^* a^* e_{-1} b w_1) \nn \\
& = &  E_0 ( a^* e_{-1} b), \ \ \ a,b \in \cao .
\ea
{}From the proof of Corollary 6 we know
$$
\{ ae_{-1} b | a,b, \in \cao \} \subset \ca_{-2,0} \ \ \mbox{is dense}.
$$
The faithfulness of $E_2$ then  proves
$$
\{E_0( ae_{-1} b)  | a,b, \in \cao \} \subset \cae
\ \ \mbox{is  dense.}
$$
Hence
\bea
\mu_{-1}:\cae^+ &\rightarrow& \bar{\bf R}^+ \cong
{\bf \bar {R}^+ 1} \subset \bar{\cae}^+ \nn \\
 x^* x & \mapsto & = \sum_i \lambda_i \gamma^{-1}(x^*x) \lambda_i^*  \nn
\ea
even defines a semifinite faithful normal weight.
To prove that $\cf$ has dense range
let $a,b \in\cao.$ Then the Plancherel formula,
 Theorem 17 and the intertwiner property of $w_1$ show
\bea
\mu_{-1} (\cf(\lambda_i)^* \cf (a)^* \cf (b)) &=&
\mu_{-1} ((\cf (a) \cf(\lambda_i) )^*  \cf (b)) =
\mu_{-1} (\cf(a * \lambda_i))^*  \cf (b)) \nn \\
&=& \mu_0 ( a*\lambda_i)^*b)
=w_1^* w_1^*  a* \gamma(\lambda_i)^*w_{-1}bw_1 \nn \\
&=&w_1^*\li^*w_1^*a^*w_{-1}bw_1.
 \nn
\ea
Therefore we get
\bea
\sum_i \mu_{-1} (\cf(\lambda_i)^* \cf (a)^* \cf (b)) \cf (\lambda_i ) & = &
\sum_i \gamma(w_{-1}^*\li w_1)\gamma(w_1^*\li^*w_1^*a^*w_{-1}bw_1) \nn \\
& = &
\gamma(\sum_i w_{-1}^*(\li w_1 w_1^*\li^*) w_1^*a^*w_{-1}bw_1) \nn \\
& = & \gamma(w_{-1}^* w_1^* a^*w_{-1}bw_1) \nn
\ea
Now $ w_{-1}^* \in \cn $ and
$ w_1^*a^* w_{-1} = \gamma^{-1}(\cf (a )^* ) \in \ca_{0,1} \subset \cn ' $ and
we conclude
$$
\hspace{2.5cm } =\gamma(w_1^* a^*w_{-1}w_{-1}^*bw_1)= \cf(a)^*\cf(b) .
$$
This shows that  the Fourier transform has dense range and, therefore,  the
continuation to an operator
$\cf :L^2(\cao,\mu_0) \rightarrow L^2(\cae, \mu_{-1})$ is a unitary.
\eot

 For the rest of this paper we will assume the depth 2 condition, i.e. the
Fourier transform has dense range.
We are now in the position to define the antipode on $\cae.$
First we get for $a,b \in \cao$
\bea
\cf(a)^*|b> & = & w_{-1}^*\gamma(a^*)w_{-3}bw_{-1}\om \nn \\
& = & w_{-1}^*\gamma(a^*)bw_{-1}w_{-1}\om =|b*a^*> \nn
\ea
where we used $b\in \cm_{-2}',w_{-3}\in \cm_{-3}\subset \cm_{-2}$ and
the intertwiner property of $w_{-1}.$
Let $S_{\cao} $ be the Tomita operator of the GNS representation
$(\cao, w_{-1}\om) .$ Then we compute
\bea
 S_{\cao}\cf(a)^*S_{\cao} |b> & = & S_{\cao} \cf(a)^* |b^*> =
S_{\cao}|b^* *a^*> \nn \\
& = & |b*a> = \cf(a^*)^*|b>.
\ea
As was proven by S. Yamagami, [Ya1], see also the  remark at the end
of this section, $\mu_0$
is a trace, i.e.
$$
S_{\cao}=J_{\cao}.
$$
Hence the Tomita operator is equal to the antiunitary modular conjugation, and
we get
$$
\mbox{Ad }J_{\cao} (\cae)=\cae .
$$
\begin{definition}
The map
\bea
\hat{S} : \cae &\rightarrow & \cae \nn \\
\hat{a}&\mapsto &\mbox{ Ad } J_{\cao} (\hat{a}^*) \nn
\ea
is called the antipode on $\cae.$ The map
$$ S:=\cf^{-1}\circ \hat{S}\circ \cf : \cao \rightarrow \cao
$$
is called the antipode on $\cao.$
\end{definition}

Using methods of S. Baaj and G. Skandalis, [Ba-Sk], the results of
section 6 imply that $\hat{S}$ really gives the antipode on $\cae.$
It then follows from the theory of Kac algebras that the antipode on
$\cao$ is given by
 $ S:=\cf^{-1}\circ \hat{S}\circ \cf .
$
Also note that equation (22) implies  $ \hat{S}(\cf(a))=\cf(a^*)^* , a
\in \cao.$

The following properties are nearly obvious.

\bigskip\bigskip
\begin{proposition}
Let $\hat{a}, \hat{b} \in \cae.$ Then
\begin{description}
\item[i)] $\hat{S}(\hat{a}\hat{b})=\hat{S}(\hat{b})\hat{S}(\hat{b})$
\item[ii)] $\hat{S}(\hat{a}^*)=\hat{S}(\hat{a})^*$
\item[iii)] $\hat{S}\circ \hat{S} = id $
\end{description}
i.e. $\hat{S}$ defines an involutive anti-automorphism on $\cae.$
Furthermore $\mu_{-1}$ is $\hat{S}$-invariant.
\end{proposition}

\proof
We only prove the $\hat{S}$-invariance of $\mu_{-1}.$
For this let $\hat{a}\in D(\mu_{-1}),$
\bea
\mu_{-1}(\hat{a})&=& \sum_i \li \hat{a}\li^*
=\sum_i <\li^*|\hat{a}|\li^*> \nn \\
& = & \sum_i <\li|J_{\cao}\hat{a}^*J_{\cao}|\li>
=\sum_i <\li|\hat{S}(\hat{a}^*)|\li>. \nn
\ea
Now $\mu_0$ is a trace and therefore $\li^*\in \cao$ is
also a Pimsner-Popa basis, i.e.
$$
= \mu_{-1}(\hat{S}(\hat{a})).
$$
\eot

It is easy to show that Proposition 19 implies the same properties for
S. Moreover, $\mu_0$ is S-invariant.

{\bf Remark}

We close this section with some remarks concerning the tracial properties
of $\mu_0$ and $\mu_{-1}.$
First note that similar to the proof of Theorem 17 one can show that
$$
E_{-2} \circ E_{-1} \circ E_0 |
 \cm \cap \gamma (\cn ') \rightarrow {\bf \bar{C}^+}
$$
defines a faithful normal weight, where the operator valued weights are
defined  by liftings.
Due to a general result of U. Haagerup, [Ha1,2], which one can check to
apply for our case, there is
 a bijection
 between semifinite normal faithful weights on
$\cm \cap \gamma (\cn ')$ and semifinite faithful normal operator valued
weights
from $ \cm \rightarrow \gamma(\cn)$, see also [St].
The  assumption of depth 2 of the underlying inclusion implies the
 factor property  of $\cm \cap \gamma (\cn ')$
 and from Corollary 6 we know $ \cm \cap \gamma (\cn ') \cong
{\cal L}(\ch),$ i.e.  there is a unique faithful normal tracial
weight on $\cm \cap \gamma (\cn )' .$
We conjecture that the associated operator valued weight from $\cm $ to
$\gamma ( \cn )$ is exactly $E_{-2} \circ E_{-1} \circ E_{0}.$
{}From the  boundedness of $E_0$ resp. $E_{-2}$ we get
$$E_{-2} \circ E_{-1} \circ E_0 | \cn \cap \gamma (\cn ') = \mu_{-1}$$
i.e. if the above conjecture holds, then  $\mu_{-1}$ is tracial.
We have been told by    S. Yamagami that he
has proven such a result,  [Ya 1,2].

In the finite index case the conjecture is true by the following arguments.
Irreducibility implies that the conditional expectations $ E_{i}$
are minimal. By the work of H. Kosaki and R. Longo,  [ Ko-Lo],
the composition of minimal conditional expectations is again minimal
and we conclude that $E_{-2} \circ E_{-1} \circ E_0 $ is minimal.
But for minimal conditional expectations
 the restriction onto the relative commutant is tracial, see  [ Lo 2].

\vspace{1cm}
In any case we have the following
\begin{lemma}
 $\mu_{-1} $ is tracial iff  $\mu_0 $ is a trace.
\end{lemma}

\proof
\noindent
Let  $\mu_0$ be a trace, $\li$
a Pimsner-Popa basis as above. Then
$$
\mu_0(\li \lambda_j^*) = \mu_0( \lambda_j^*\li ) = \delta_{i,j},
$$
i.e. $\li^*$ is again a Pimsner-Popa basis. We get for $x \in \cae$
\bea
\mu_{-1}(\gamma(x^*x)) & = & \sum_i \li^* x^*x \li =
\sum_i \li x^*x \li^* \nn \\
& = & w_{-1}^* \sum_i \li^* x^*x \li w_{-1} =
\sum_i <\li | x^*x| \li > \nn \\
& = & \mbox{tr}_{\ch} x^*x .
\ea
This also proves $\mu_{-1}$ tracial.

To prove the opposite direction let us note a general result.
Define
\bea \hat{\mu_0} : \cao^+ & \rightarrow & \overline{\ca_{-3,-1}}^+ \nn \\
a & \mapsto & \sum \cf(\li) \gamma( a) \cf (\li)^* .\nn
\ea
Then we compute for $a\in \cao^+$
\bea
\hat{\mu_0}(a) & = & \gamma(\sum_i w_{-1}^* \li w_1 a w_1^*\li^*w_{-1}) \nn \\
& = & \gamma(\sum_i w_{-1}^*\gamma(a) \li w_1  w_1^*\li^*w_{-1}) \nn
\ea
where we used the intertwiner property of $w_1$ and $ \li \in \gamma(\cm)'.$
{}From $\li$ being a Pimsner-Popa basis for $E_0:\cm\rightarrow \cn$
we conclude
$$
\hspace{1cm} = \gamma(w_{-1}^* \gamma(a) w_{-1}) = \mu_0(a).
$$

Using the Plancherel formula, Theorem 17, we get
$$
\cf(\li) \in \cae \ \ \mbox{ is an orthonormal basis in }
\ L^2(\cae, \mu_{-1}).
$$

Now the definition of $\hat{\mu_0}$ does not depend on the
special choice of such
a basis and we can argue as before, i.e. if $\mu_{-1} $ is tracial,
we get $\mu_0$ is a trace.

\eot

\section{The Crossed Product}

In this section we identify
$$\ca_{0,1} \cong \cae =\gamma(\ca_{0,1})$$
and put
$$\cf_1 := \gamma^{-1} \circ \cf, \ \ \ \hat{S}_1 := \gamma^{-1}\circ
\hat{S} \circ \gamma ,
$$
and

$$
\hat{\varepsilon}_1 =\hat{\varepsilon} \circ \gamma.
$$
We then extend the natural right action of the compact
Kac algebra $\ca_{0,1}$ on its dual $\cao \subset \cm$ to
an action on the whole of $\cm,$ such that $\cn \subset \cm$ is given
as the $\ca_{0,1}$ invariant subalgebra and the conditional
expectation $E_0: \cm \rightarrow \cn$ is given as the right action of the
integral ( $\equiv$ Haar element ) $e_{1} \in \ca_{0,1}.$
We then prove that $\cm_1 $ is in fact the crossed product
$$
\cm_1 = \cm \lhd \ca_{0,1} . $$
Throughout this section we assume the depth 2 condition on $\cn \subset \cm.$

\begin{definition}
For $a\in \cao$ and $m\in \cm$ we put
$$
a*m := w_{-1}^*a\gamma(m)w_{-1}.
$$
\end{definition}
Note that for $m\in \cao \subset \cm$ this coincides with our previous
convolution product, equ. (11).
Moreover we have
\begin{proposition}
Let $a,b \in \cao$ and $m\in\cm$.
$$
i) \ \ \ (a*b)*m=a*(b*m)
$$
$$
ii)\ \ \ {\bf 1} *m = E_0(m)
$$
\hspace{3cm} iii) The following conditions a) - d) are equivalent
\begin{description}
\item[a)] $n \in \cn$
\item[b)] $a*n=\mu_0(a)n \ \ \forall a \in \cao$
\item[c)] $a*(nm)=n(a*m), \ \forall a\in\cao, m\in \cm$
\item[d)] $a*(mn)=(a*m)n, \ \forall a\in\cao, m\in \cm$
\end{description}
\end{proposition}

\proof

\noindent
i) is straight forward from the definitions.

\noindent ii)
$
{\bf 1}*m = w_{-1}^* \gamma (m) w_{-1} = E_0(m)
$
by Lemma 2.

\noindent iii) Let us show the equivalence.
a) $\Rightarrow$ b) :
$a*n=w_{-1}^*a\gamma(n)w_{-1}=w_{-1}aw_{-1}n= \mu_0(a) n$
where we used the intertwiner property of $w_{-1}$ and Corollary 3.

a) $\Rightarrow$ c) :
$
a*(nm) = w_{-1}^*\gamma(nm)a w_{-1}=
 n w_{-1}^*\gamma(m)a w_{-1}  = n(a*m) .
$

a) $\Rightarrow$ d) is similarly proved as a) $\Rightarrow$ c), using
$\gamma(n)a=a\gamma(n)$.

b) $\Rightarrow$ a) : Put $a = {\bf 1}$ and apply ii)

c) $\Rightarrow$ b) and d) $\Rightarrow$ b) :
 Put $m={\bf 1}$ and note that $a*{\bf 1} = w_{-1}^*aw_{-1}=\mu_0(a),$
see Corollary 3.
\eot

We now use Proposition 22 to construct a right Hopf-module action of
$\ca_{0,1}$ on $\cm$ which extends the natural right action of
$\ca_{0,1}$ on its dual $\cao \subset \cm$ and which becomes an inner
action in $\cm_1.$ In order to avoid analytical problems at this
point, let us assume for the moment finite index. It will follow from the
results of
section 6 that the formulas
also hold in the infinite index case.

\begin{definition}
For $\hat{a}\in \ca_{0,1}$ and $m\in \cm$ denote
$$
m \lhd \hat{a} := \cf_1^{-1}(\hat{S}_1(\hat{a}))*m \ \ \in \cm
$$
\end{definition}
This will be our right action.
To have explicit formulas we put $w_{2n}:=\gamma^{-n}(w_0)$ and
compute   for $\hat{a}\in \ca_{0,1}, m\in \cm$
$$
\cf_1^{-1}(\hat{a})=w_0^*\gamma(\hat{a})w_{-2}
$$
$$
\cf_1^{-1}(\hat{S}_1(\hat{a})) = \cf_1^{-1}(\hat{a}^*)^*= w_{-2}^*
\gamma(\hat{a})w_0 \ \ \ \in
\cao
$$
and therefore
\bea
m \lhd \hat{a}& = &
w_{-1}^*\cf_1^{-1}(\hat{S}_1(\hat{a}))\gamma(m)w_{-1} \nn \\
& =& w_{-1}^*\gamma(m)\cf_1^{-1}(\hat{S}_1(\hat{a}))w_{-1} \nn \\
&=& w_{-1}^*\gamma(m)w_{-2}^*\gamma(\hat{a})w_0w_{-1} .
\ea
Now we use the intertwiner property (15) for $w_0$, which
 together with $\gamma(\hat{a})\in \cae\subset \cm_3'$ yields
\bea
\hspace{1cm}&=&
w_{-1}^*\gamma(m)w_{-2}^*w_{-3}\gamma(\hat{a})w_0 \nn \\
&=&w_{-1}^*\gamma(m)\gamma(\hat{a})w_0
\ea
where in the last line we used  equation (16).

This later formula also makes sense in the infinite index case.

\bigskip
To prepare Theorem 26 below we need the following two Lemmas.
\begin{lemma}
For $m\in \cm$ and $a\in \cao$ we have
$$
(m\lhd \cf_1(a))^*=m^*\lhd\cf_1(a^*)
$$
\end{lemma}

\proof
Using the antipode on $\cao,\ \  S=\cf_1^{-1}\circ \hat{S}_1 \circ \cf_1,$
and
Lemma 12 we have
\bea
(m\lhd \cf_1(a))^* & = & (S(a)*m)^* = S(a)^* * m^* \nn \\
& = & m^* \lhd \cf_1(a)^*. \nn
\ea
Here we used that $S$ commutes with $*,$ which follows from
Proposition 19 ii),iii) and the identity $\cf_1 (a^*)
=\hat{S}_1(\cf_1(a))^* $ implying
\bea
^* \circ S \circ ^* &=& (\cf_1 \circ ^*)^{-1}\circ \hat{S}_1 \circ
(\cf_1 \circ ^* )=(^* \circ \hat{S}_1 \circ \cf_1)^{-1} \circ
\hat{S}_1 \circ (^* \circ \hat{S}_1 \circ \cf_1) \nn \\
& = & \cf_1^{-1} \circ \hat{S}_1^3 \circ \cf_1 = S \nn
\ea
\eot

\begin{lemma}
For $a,b\in \cao$ we have
$$
\cf_1(b\lhd \cf_1(a))=\cf_1(a^*)^*\cf_1(b)
$$
\end{lemma}

\proof
As in the above proof we compute
\bea
\cf_1 (b \lhd \cf_1(a))& = & \cf_1(S(a)*b)=\cf_1(S(a))\cf_1(b) \nn \\
& = & \hat{S}_1 (\cf_1(a))\cf_1(b) = \cf_1(a^*)^* \cf_1(b). \nn
\ea

\eot

Now we are in the position to formulate

\begin{theorem}
For $\hat{a},\hat{b} \in \ca_{0,1}$ and $m \in \cm$ we have
\begin{description}
\item[i)]$(m\lhd \hat{a})\lhd \hat{b}= m\lhd(\hat{a}\hat{b}) $
\item[ii)]$m\lhd e_1=E_0(m)$
\item[iii)]$m\in \cn \Leftrightarrow m\lhd \hat{a}
=\hat{\varepsilon}_1(\hat{a})m \ \ \forall \hat{a}\in \ca_{0,1} $
\item[iv)]$m\in \ca_{-1,0}\Rightarrow m\lhd
\hat{a}=<m^{(1)},\gamma(\hat{a})>m^{(2)} $
\item[v)]In $\cm_1$ we have
$$
m\lhd \hat{a}=\hat{S}_1(\hat{a}^{(1)})m\hat{a}^{(2)}
$$
where $\hat{a}^{(1)}\otimes \hat{a}^{(2)}=\hat{\Delta}(\hat{a})$
is the natural coproduct on $\ca_{0,1}$ obtained from the pairing with
$\cao.$
\end{description}
\end{theorem}

\proof
i) follows from Proposition 22 i), Proposition 19 i) and Lemma 9.

\noindent ii) follows from Proposition 22 ii), since $\cf_1({\bf 1})=e_1.$

\noindent iii) follows from Proposition 22 iii), since
$\mu_0(a)=\hat{\varepsilon}_1 (\cf_1(a)).$

\noindent iv) Let $\hat{a}=\cf_1(a)$ and $\hat{b}=\cf_1(b).$ Then iv)
is equivalent to
\be
<(m\lhd \hat{a}), \gamma(\hat{b})>=<m,\gamma(\hat{a}\hat{b})>\equiv
\mu_0(m(a*b)), \ \forall m \in \cm, a,b \in \cao
\ee
To prove  (25) we use the  Lemmas 25 and 26 and the Plancherel
formula, Theorem 17, to get
\bea
<(m\lhd \hat{a}), \gamma(\hat{b})> & = & \mu_0((m\lhd \hat{a})b) \nn \\
& = & \mu_1(\cf_1(m^*\lhd\cf_1(a^*))^*\hat{b}) \nn \\
&=& \mu_1(\cf_1(m^*)^* \cf_1(a)\cf_1(b))) = \mu_0(m(a*b)) . \nn
\ea
v) To prove v) we use the identity (24), $m \lhd \hat{a}=w_{-1}^*
\gamma(m\hat{a})w_0 ,$ and the fact that also $S(\li)$ is a
Pimsner-Popa basis, ( since S extends to an unitary ), to compute
\bea
m \lhd \hat{a} & = & \sum_i w_{-1}^* S(\li) w_1 w_1^* S(\li)^*
\gamma(m)\gamma(\hat{a})w_0 \nn \\
& = & \sum_i \cf_1(S(\li))m (S(\li)w_1)^* \gamma(\hat{a})w_0 \nn \\
& = & \sum_i \hat{S}_1 (\cf_1(\li)) m
w_0^*\cf_1(\li^*)\gamma(\hat{a})w_0 \nn
\ea
where we have used $S= \cf_1^{-1} \circ \hat{S}_1 \circ \cf_1$ and
therefore
$$
S(\li)w_1= \cf_1^{-1} (\cf_1(\li^*)^*)w_1 = \cf_1(\li^*)^*w_0 .
$$
Now $\cf_1(\li^*) \in \ca_{0,1}$ commutes with $\gamma(\hat{a})\in \cae
\subset \cn$ and $ \cf_1(\li^*)w_0=\li^* w_1 .$ Hence v) holds,
provided
$$
\hat{\Delta}(\hat{a})=\sum_i \cf_1(\li)\otimes
w_0^*\gamma(\hat{a})\li^* w_1.
$$
To check this formula for $\hat{\Delta}$ we have to use the pairing
between $\cao$ and $\cae$ and verify that $\hat{\Delta}(\hat{a}) $ is
the solution of
$$
\mu_0(bc\cf_1^{-1}(\hat{a}))=(\mu_0 \otimes \mu_0 )
(( b\otimes c) (\cf_1^{-1} \otimes \cf_1^{-1}
)(\hat{\Delta}(\hat{a})))
\ \ \ \forall b,c \in \cao.
$$
Using the above expression for $\hat{\Delta}(\hat{a}) ,$
the r.h.s. gives
\bea
(\mu_0 \otimes \mu_0 )
(( b\otimes c) (\cf_1^{-1} \otimes \cf_1^{-1}
)(\hat{\Delta}(\hat{a}))) & = & \sum_i \mu_0(b\li)\mu_0(c
\cf_1^{-1}(w_0^*\gamma(\hat{a})\li^* w_1)) \nn \\
& = &
\mu_0(c\cf_1^{-1}(w_0^*\gamma(\hat{a})b w_1)), \nn
\ea
where we used that $\li$ is a Pimsner-Popa basis for
$\mu_0:\cao \rightarrow {\bf C}.$
Now
\bea
w_0^* \gamma(\hat{a}) bw_1 & = & w_0^* \gamma(\hat{a})w_{-3}^* w_{-2}
b w_1 \nn \\
& = &
w_{-1}^* w_0^* \gamma(\hat{a})w_{-2}bw_1 \nn \\
& = &
\cf_1(\cf_1^{-1}(\hat{a})b). \nn
\ea
Hence, using the trace property of $\mu_0$ we get
$$
(\mu_0 \otimes \mu_0 )
(( b\otimes c) (\cf_1^{-1} \otimes \cf_1^{-1}
)(\hat{\Delta}(\hat{a})))=\mu_0(bc\cf_1^{-1}(\hat{a})),
$$
i.e. the defining equation for $\hat{\Delta}(\hat{a}).$
This concludes the proof of v) and hence of the Theorem 26.
\eot

The statements iii) and iv) of Theorem 26 imply
$$
\cn = \ca_{0,1}' \cap \cm_1
$$
and since clearly
$$
\cm_1 = \cm \vee \ca_{0,1}
$$
we have the
\begin{corollary}
$\ca_{0,1}$ acts outerly on $\cm$ and $\cm_1 = \cm \lhd \ca_{0,1}$ is
a crossed product.
\end{corollary}

\section{The Multiplicative Unitary W}
In this section we finish our analysis showing that the
coproduct$\Delta$ is actually a *-algebra homomorphism $\Delta:\cao
\rightarrow \cao \otimes \cao $ such that $(\cao, \Delta , {S},\mu_0)$ becomes
a discrete Kac algebra with antipode ${S}$  and Haar state
$\mu_0.$ We also verify $\cae$ to be the dual compact Kac algebra and
rederive the pairing formula. Our methods will heavily rely on the
results of S. Baaj and G. Skandalis, [Ba-Sk], who used the notion of a
multiplicative unitary W as a basic tool. Given a $C^*$-Hopf algebra
$\ca$ with coproduct $\Delta$ and Haar state $\mu_0$ the operator
W is defined on
$L^2(\ca,\mu_0)
\otimes L^2(\ca,\mu_0)$ by
$$
W|a\otimes b> := \Delta (a) |{\bf 1} \otimes b>, \ \ \ a,b \in \ca .
$$
Then W is unitary and satisfies the pentagon identity
\be
W_{23}W_{12}=W_{12}W_{13}W_{23}.
\ee
Conversely, given a unitary
$$
W:\ch \otimes \ch \rightarrow \ch \otimes \ch
$$
on some Hilbert space $\ch,$ such that W satisfies the above pentagon
identity (26) together with some regularity conditions, S. Baaj and
G. Skandalis, [Ba-Sk], have shown how to recover in ${\cal L}(\ch)$ an
underlying Hopf algebra $\ca$ and its dual $\hat{\ca},$
such that W is given by the above formula.

This is the route we will follow. We first use our definition of
$\Delta$, equation (12), to define a candidate for W in terms of its
matrix elements
$$<\lambda_k\otimes \lambda_l|W(\lambda_i\otimes \lambda_j)>
:=
(\mu_0 \otimes \mu_0)
((\lambda_k^*\otimes \lambda_l^*)\Delta(\li)({\bf 1}\otimes \lambda_j)).
$$
Now we use the trace property of $\mu_0$ to get
$$=
(\mu_0 \otimes \mu_0)
((\lambda_k^*\otimes \lambda_j \lambda_l^*)\Delta(\li)),
$$
and therefore by the definition of $\Delta$
$$
= \mu_0((\lambda_k^**(\lambda_j\lambda_l^*))\li).
$$
We will then show that W is indeed a multiplicative unitary and that
the reconstruction procedure of [Ba-Sk] precisely reproduces $\ca \equiv
\cao$ and $\hat{\ca}\equiv \cae$ together with the pairing
leading to our coproduct $\Delta.$ We also verify that $\hat{S}$ given
in Definition 18 is indeed the antipode.

To simplify computations we first consider
$\hat{W} := $Ad $J_{\cao} (W)$ which gives
\bea
<\lambda_k\otimes\lambda_l|\hat{W}(\lambda_i\otimes\lambda_j)>&:=&
(\mu_0 \otimes \mu_0)
((\lambda_k^*\otimes \lambda_l^*)({\bf 1}\otimes
\lambda_j)\Delta(\li)) \nn \\
& =&
\mu_0((\lambda_k^**(\lambda_l^*\lambda_j))\li)
\ea
where again we used the trace property of $\mu_0,$ i.e. the fact that $\li^*$
is also a Pimsner-Popa basis.
Using Corollary 3, a simple computation shows that $\hat{W}$ is given
on $\ch \otimes \ch$ by
\be
\hat{W} |\lambda_i \otimes \lambda_j>=\sum_{k,l} z^{kl}_{ij}|\lambda_k
\otimes \lambda_l>
\ee
where $z^{kl}_{ij}\in \cm_1 \cap \cm' = {\bf C}$ is,
\be
z^{kl}_{ij} = w_1^*  w_1^* w_1^*  \lambda_k^* \gamma(\lambda_l^*
\lambda_j) w_{-1} \li w_1 w_1 .
\ee
The above sum converges and we actually have

\begin{theorem}
\bea
\hat{W} : \ch \otimes \ch & \rightarrow & \ch \otimes \ch \nn \\
|\lambda_i > \otimes |\lambda_j > & \mapsto &
\sum_{k,l}
(w_{1}^* w_{1}^*w_{1}^*\lambda_k^* \gamma(\lambda_l^* \lj) w_{-1} \li w_1 w_1)
\ |\lambda_k > \otimes |\lambda_l> \nn
\ea
defines a multiplicative unitary matrix.
\end{theorem}

\proof
{\bf i) Unitarity}
\bea
\| \hat{W}(a\otimes b) \|^2 & = &
\sum_{k,l}
(w_{1}^* w_{1}^*w_{1}^*\lambda_k^* \gamma(\lambda_l^* b) w_{-1} a w_1 w_1)^*
(w_{1}^* w_{1}^*w_{1}^*\lambda_k^* \gamma(\lambda_l^* b) w_{-1} a w_1 w_1)
\nn \\
& = &
\sum_{k,l}
(w_{1}^* w_{1}^*a^* w_{-1}^* \gamma(b^*\lambda_l)\lambda_k w_{1} w_1 w_1)
(w_{1}^* w_{1}^*w_{1}^*\lambda_k^* \gamma(\lambda_l^* b) w_{-1} a w_1 w_1)
\nn \\
& = &
\sum_{k,l}
(w_{1}^* w_{1}^*a^* w_{-1}^* \gamma(b^* )\lambda_k (\gamma(\lambda_l) w_{1} w_1
w_1)
(w_{1}^* w_{1}^*w_{1}^* \gamma(\lambda_l^*)) \lambda_k^* \gamma( b) w_{-1} a
w_1 w_1 \nn
\ea
Using the intertwiner property of $w_1,$   i.e.
$$  (\gamma(\lambda_l) w_{1} w_1 w_1 ) =
 w_1 \lambda_l w_1 w_1 $$
and Corollary 4, i.e.
$$
w_{1}^* w_{1}^*a^* w_{-1}^* \gamma(b^* )\lambda_k w_1 \lambda_l w_1 =
c w_1^* ,\ \ \ c \in {\bf C},
$$
some of the factors $w_1^*, w_1$ cancel due to
$ w_1^*w_1={\bf 1},$ see Corollary 1, and we get
$$
\| \hat{W}(a\otimes b) \|^2  =
\sum_{k,l}
w_{1}^* w_{1}^*a^* w_{-1}^* \gamma(b^* )\lambda_k
w_1 \lambda_l  w_1
w_{1}^* \lambda_l^*w_1^* \lambda_k^*
 \gamma( b) w_{-1} a w_1 w_1. $$
$\li$ being a Pimsner-Popa basis implies
\bea
\hspace{3cm} &=& w_{1}^* w_{1}^*a^* w_{-1}^* \gamma(b^* )
 \gamma( b) w_{-1} a w_1 w_1 \nn \\
& = & w_{-1}^* \gamma(b^* )
 \gamma( b) w_{-1} w_1^* a^* a w_1 \nn \\
 & = & \mu_0 ( b^* b) \mu_0(a^* a)=\|a\otimes b\|^2 \nn
\ea
by Corollary 3.
Similarly one proves
$$ \| \hat{W}^*(a\otimes b) \|^2 = \mu_0 ( b^* b) \mu_0(a^* a) .$$

{\bf ii)  Pentagon Relation}
Using (28), the Pentagon relation (26) reads as follows,
\be
z_{j_2,i_3}^{n_2,n_3} z_{i_1,i_2}^{n_1,j_2} =
z_{j_1,j_2}^{n_1,n_2} z_{i_1,j_3}^{j_1,n_3} z_{i_2,i_3}^{j_2,j_3}
\ee
where the summation is taken over doubled indices.
We compute the l.h.s. of (30)
$$  \ \  z_{j_2,i_3}^{n_2,n_3} z_{i_1,i_2}^{n_1,i_2}  $$
$$=
\sum_{j_2}
w_{1}^* w_{1}^*w_{1}^*\lambda_{n_2}^*
\gamma(\lambda_{n_3}^* \lambda_{i_3}) w_{-1} \lambda_{j_2} w_1 w_1
w_{1}^* w_{1}^*w_{1}^*\lambda_{n_1}^*
\gamma(\lambda_{j_2}^* \lambda_{i_2}) w_{-1} \lambda_{i_1} w_1 w_1
$$
$$=
\sum_{j_2}
w_{1}^* w_{1}^*w_{1}^*\lambda_{n_2}^*
\gamma(\lambda_{n_3}^* \lambda_{i_3}) w_{-1} \lambda_{j_2} w_1 w_1
w_{1}^* w_{1}^*w_{1}^* \gamma(\lambda_{j_2}^*)
\gamma( \lambda_{i_2}) \lambda_{n_1}^* w_{-1} \lambda_{i_1} w_1 w_1.
$$
Using the intertwiner property of $w_1,$ $ w_{1}^* \gamma(\lambda_{j_2}^*)
= \lambda_{j_2}^* w_1^*$ and applying Corollary 4 as in the above
proof of unitarity,
we get similarly
$$=
w_{1}^* w_{1}^*w_{1}^*\lambda_{n_2}^*
\gamma(\lambda_{n_3}^* \lambda_{i_3}) w_{-1} w_1^*
\gamma( \lambda_{i_2}) \lambda_{n_1}^* w_{-1} \lambda_{i_1} w_1 w_1.
$$
We shift $w_1^*$ from the middle to the left using
 the intertwining property which leads to
$$=
w_1^* w_{1}^* w_{1}^*w_{1}^*\lambda_{n_1}^* \gamma (\lambda_{n_2}^*)
\gamma^2(\lambda_{n_3}^* ) \gamma^2(\lambda_{i_3}) w_{-3}
\gamma( \lambda_{i_2})  w_{-1} \lambda_{i_1} w_1 w_1,
$$
where we  made use of $[\gamma(a), \lambda^*_{n_1}] =0 $ for all $a
\in \cm.$

For the r.h.s. of (30) we first compute
$$  \ \  z_{j_1,j_2}^{n_1,n_2} z_{i_1,j_3}^{j_1,n_3}  $$
$$=
\sum_{j_1}
w_{1}^* w_{1}^*w_{1}^*\lambda_{n_1}^*
\gamma(\lambda_{n_2}^* \lambda_{j_2}) w_{-1} \lambda_{j_1} w_1 w_1
w_{1}^* w_{1}^*w_{1}^*\lambda_{j_1}^*
\gamma(\lambda_{n_3}^* \lambda_{j_3}) w_{-1} \lambda_{i_1} w_1 w_1.
$$
Using similar arguments as before  we get
$$=
w_{1}^* w_{1}^*w_{1}^*\lambda_{n_1}^*
\gamma(\lambda_{n_2}^* \lambda_{j_2}) w_{-1 }w_{-3}^*
\gamma(\lambda_{n_3}^* \lambda_{j_3}) w_{-1} \lambda_{i_1} w_1 w_1.
$$
Now we push $w_{-3}^*$ to the left. Using
$ w_{-1} w_{-3}^* = w_{-5}^* w_{-1}$ and  $(w_{1}^*)^3
w_{-5}^*=(w_1^*)^4$  we get
\be =
w_{1}^* w_{1}^*w_{1}^*w_1^* \lambda_{n_1}^*
\gamma(\lambda_{n_2}^* ) \gamma^2(\lambda_{n_3}^*)
\gamma(\lambda_{j_2}) w_{-1} \gamma(\lambda_{j_3}) w_{-1}
\lambda_{i_1} w_1 w_1 . \nn
\ee

Finally we compute
$z^{j_2,j_3}_{i_2,i_3}=w_1^*\tilde{z}^{j_2,j_3}_{i_2,i_3} w_1$ where
\bea
 \ \ \ \tilde{z}_{i_2,i_3}^{j_2,j_3}
& = & w_1^*w_1^*\lambda_{j_2}^*\gamma(\lambda_{j_3}^*\lambda_{i_3})
w_{-1}\lambda_{i_2} w_1 \nn \\
& = &
  w_{1}^*w_{1}^*\gamma( \lambda_{j_3}^* )
\lambda_{j_2}^* \gamma(\lambda_{i_3}) w_{-1} \lambda_{i_2} w_1 \nn \\
& = & w_{1}^* \lambda_{j_3}^* w_1^*
\lambda_{j_2}^* \gamma(\lambda_{i_3}) w_{-1} \lambda_{i_2} w_1 \in
{\bf C}\nn \\
\ea
where we used the intertwiner property of $w_1.$
Now $\tilde{z}_{i_2,i_3}^{j_2,j_3}\in \ca_{1,1}={\bf C}$ and hence
$ z_{i_2,i_3}^{j_2,j_3}=\tilde{z}_{i_2,i_3}^{j_2,j_3}
= \gamma(\tilde{z}_{i_2,i_3}^{j_2,j_3})$ and we conclude
$$
z_{i_2,i_3}^{j_2,j_3}= w_{-1}^* \gamma(\lambda_{j_3}^*) w_{-1}^*
\gamma(\lambda_{j_2}^*) \gamma^2(\lambda_{i_3}) w_{-3}
\gamma(\lambda_{i_2}) w_{-1}.
$$
Plugging this expression into the formula (32) between
$\gamma(\lambda_{j_3}) w_{-1}$ and $ \lambda_{i_1} $ and using twice
 the property of $ \li$ being a
 Pimsner-Popa-basis  we conclude
\bea
z_{j_1,j_2}^{n_1,n_2} z_{i_1,j_3}^{j_1,n_3} z_{i_2,i_3}^{j_2,j_3}
& = &
w_1^*w_{1}^* w_{1}^*w_{1}^*\lambda_{n_1}^* \gamma (\lambda_{n_2}^*)
\gamma^2(\lambda_{n_3}^* ) \gamma^2(\lambda_{i_3}) w_{-3}
\gamma( \lambda_{i_2})  w_{-1} \lambda_{i_1} w_1 w_1 \nn \\
& = & z_{j_2,i_3}^{n_2,n_3} z_{i_1,i_2}^{n_1,j_2} \nn
\ea
which proves the Pentagon identity.

\hfill $\Box$

\noindent Let us remark here that the definition of the multiplicative matrix
$\hat{W}$ and its properties do not rely on the tracial
property of $\mu_0.$

\noindent We get as a simple

\begin{corollary}

$W:= \mbox{Ad }J_{\cao}(\hat{W})$ is a multiplicative unitary.

\end{corollary}

We now use the reconstruction theorems of S. Baaj and G. Skandalis,
[Ba-Sk], to show that $\cae$ and $\cao$ are indeed the pair of dual Hopf
algebras associated with W. To this end we first perform the
reconstruction starting from $\hat{W}$ and then use $W=\mbox{Ad }
J_{\cao} (W)$ to arrive at our final results.

For  $a,b \in \cao,$ let
$$
\omega_{a,b}:= <a| \cdot |b>, \ \ \in (\cao)_* \ \ .
$$
Following [Ba-Sk] we define
\bea
\hat{\rho}(\omega_{a,b})&:=&({\bf 1} \otimes \omega_{a,b})(\hat{W}) \nn \\
\hat{L}(\omega_{a,b})&:=& (\omega_{a,b} \otimes {\bf 1})(\hat{W}). \nn
\ea
The statement of S. Baaj and G. Skandalis , [Ba-Sk], is that the *-closure of
$$
\{ \hat{\rho}(\omega_{a,b}) \in {\cal L}(\ch) | a,b \in \cao\}
$$
generate a Hopf-algebra with dual generated by
$$
\{ \hat{L}(\omega_{a,b}) \in {\cal L}(\ch) | a,b \in \cao\} .
$$

We compute

\begin{lemma}
For $a,b\in \cao$ we have
$$
i) \ \  \hat{\rho}(\omega_{a,b}) = \cf(a^*b), \ \ a,b \in \cao,
$$
\bea
ii) \ \ \ \hat{L}(\omega_{a,b})&=&\mbox{Ad }J_{\cao}
(\sum_i \li w_{-1}^*\gamma(\li)^*
w_{-1}^*\gamma(a)^*w_{-3}\gamma(b)w_{-1})^*
\in \cao' \cap {\cal L}(\ch) \nn \\
& & \ \ \ \ \ \mbox{ for } a,b \in \cao . \nn
\ea
\end{lemma}

\proof
i) Let $a,b \in \cao .$ We get from the definition
\bea
\hat{\rho}(\omega_{a,b}) & = &
\sum_{i,k} w_{-1}^*  w_{-1}^* \lambda_k^*\gamma(a^*b)w_{-1}\lambda_i w_{-1}
|\lambda_k><\lambda_i| \nn \\
& = &
\sum_{i,k} w_{-1}^*  \lambda_k^*(w_{-3}^* \gamma(a^*b)w_{-1})\lambda_i w_{-1}
|\lambda_k><\lambda_i| \nn \\
& = & w_{-3}^* \gamma(a^*b)w_{-1}= \cf(a^*b) \in {\cal L} (\ch). \nn
\ea

\noindent
ii) A similar computation as in i) shows
\bea
\hat{L}(\omega_{a,b}) & = &
\sum_{j,l} w_{1}^*  w_{1}^* \gamma(\lambda_l^*\lambda_j)a^* w_{-1} b w_{1}
|\lambda_l><\lambda_j| \nn \\
& = &\sum_{j,l} w_{1}^*   \lambda_l^*\lambda_j w_{1}^* a^*w_{-1} b w_{1}
|\lambda_l><\lambda_j| .\nn
\ea
Now we use equality (16).
 Then we can rewrite
\bea
\hspace{1cm}& = &
\sum_{j,l} w_{1}^*   \lambda_l^*\lambda_j w_0^*w_1 w_{1}^* a^*w_{-1} b w_{1}
|\lambda_l><\lambda_j| \nn \\
& = &
\sum_{j,l} w_{1}^*   \lambda_l^*\lambda_j (w_0^* w_{-1}^* \gamma(a^*)
w_{-3} \gamma(b) w_{-1})w_1
|\lambda_l><\lambda_j| \nn
\ea
One easily shows
$$
(w_0^* w_{-1}^* \gamma(a^*)
w_{-3} \gamma(b) w_{-1}) \in \cm \cap \gamma(\cm)'= \cao ,
$$
see for example the computation in (10),
and we get
\bea
\hat{L}(\omega_{a,b}) & = &
\sum_{i} |\lambda_i(w_{0}^*  w_{-1}^* \gamma(a^*) w_{-3} \gamma(b) w_{-1} )>
<\lambda_i| \nn \\
& = &
\mbox{Ad }J_{\cao}
(w_{0}^*  w_{-1}^* \gamma(a^*) w_{-3} \gamma(b) w_{-1})^*, \nn
\ea
which proves ii) .
\eot

In order to recover our setting we now have to take the multiplicative
unitary $W$ instead of $\hat{W}.$
This gives
\bea
\rho(\omega_{a,b})= J_{\cao} \hat{\rho}(\omega_{a^*,b^*}) J_{\cao} \nn \\
L(\omega_{a,b})= J_{\cao} \hat{L}(\omega_{a^*,b^*}) J_{\cao} \nn
\ea
from which we conclude
\bea
\rho(\omega_{a,b})&=&  \cf(ba^*) \in \cae \\
L(\omega_{a,b})&=&
\sum_i  w_{-1}^*\gamma(b)w_{-3}^*\gamma(a)^*w_{-1}\gamma(\li) w_{-1} \li^*
\in \cao .
\ea
It is clear that the $C^*$-closures generated by these operators give
$\cae$ and $\cao,$ respectively.

Next,  according to [Ba-Sk] the
 natural pairing between the two algebras is given by
$$
<L(\omega_{a,b})|\rho(\omega_{c,d})>=<a\otimes c |W|b\otimes
d>=\mu_0(c^*L(\omega_{a,b})d)
$$
which by the trace property gives
\bea
\hspace{1cm}& = & \mu_0(dc^*L(\omega_{a,b})) \nn \\
& = & \mu_{0}(\cf^{-1}(\rho(\omega_{c,d}))L(\omega_{a,b}))
\ea
i.e. the natural pairing of our setting.

Finally we
 look at the antipode on $\cae$ as defined by [Ba-Sk],
$$
S_{B-S}(\rho(\omega_{c,d})):= ({\bf 1}\otimes \omega_{c,d})(W^*).
$$
Computing the r.h.s. similarly as before we get
$$
S_{B-S}(\cf(ba^*))= \cf(ab^*)^*,
$$
which due to Definition 18 and equation (22) coincides with our antipode
$\hat{S}.$

\bigskip
We conclude by mentioning without proof that we could have chosen to
work with the opposite coproduct $\Delta_{op}$ on $\cao$. This would give
$W_{op}$ and $\hat{W}_{op}$, respectively.
Putting $R=\hat{W}_{op} \sigma,$ where $\sigma$ denotes the flip on
$\ch \otimes \ch,$ we recover Longo's definition, [Lo1], of the
R-operator of J. Cuntz, [Cu].

\bigskip\noindent
{\bf Acknowledgements }

\noindent
Besides giving our gratitudes to B. Schroer for many stimulating
discussions
we would like to thank D. Bisch, R. Longo,  E. Kirchberg and K.-H. Rehren
for all the comments and helps they gave to us.


\newpage

\begin{thebibliography}{12345678}
\bibitem[Ba-Sk]{[Ba-Sk]} S. Baaj, G. Skandalis
 ``Unitaires Multiplicatifs et Dualit\'e pour les produits croises
de $C^*$-alg\'ebres''
Ann. Sci. ENS {\bf 26} (1993) p.425
\bibitem[B-D-H]{[B-D-H]} M. Baillet, Y. Denizeau, J.-F. Havet
 ``Indice d'une esperance conditionell '' Comp. Math. {\bf 66} (1988) p.199
\bibitem[Cu]{[Cu]} J. Cuntz ``Regular Actions of Hopf algebras on the
$C^*$-algebra generated by a Hilbert space'' preprint
\bibitem[Da]{[Da]} M.-C. David ``Paragroupe d'Adrian Ocneanu et algebre
de Kac'' preprint
\bibitem[Do-Ro]{[Do-Ro]} S. Doplicher, J. Roberts
 ``Why there is a field algebra with a compact gauge group
describing the superselection structure in particle physics",
Comm. Math. Phys. {\bf 131} (1990) p. 51
\bibitem[E-N]{[E-N]} M. Enock, R. Nest ``Irreducible Inclusions
of Factors and Multiplicative Unitaries'' preprint March '94
\bibitem[E-S]{[E-S]} M. Enock, Schwarz ``Kac algebras and Duality''
Springer Verlag (1992)
\bibitem[He-Oc]{[He-Oc]} R. Herrmann, A. Ocneanu
``Index Theory and Galois Theory for Infinite Inclusions of Factors''
C.R. Acad. Sci. Paris {\bf 309} (1989) p.923
\bibitem[Haa]{[Haa]} R. Haag ``Local Quantum Physics", Springer (1992).
\bibitem[Ha1,2]{[Ha1,2]} U. Haagerup ``Operator
valued weights in von Neumann algebras I,II''
J. Func. An {\bf 32 } (1979) p. 175, J. Func. An {\bf 33} (1979) p.119
\bibitem[Jo]{[Jo]} V.F.R. Jones ``Index of Subfactors'' Invent. Math.
{\bf 72} (1983) p.1
\bibitem[Ko]{[Ko]} H. Kosaki ``Extension of Jones theory on index to arbitrary
factors'' J. Funct. An. {\bf 66} (1986) p.123
\bibitem[Ko-Lo]{[Ko-Lo]} H. Kosaki, R. Longo ``A Remark on the minimal
index of subfactors'' J. Func. An. {\bf 107} (1992) p.458
\bibitem[Lo1]{[Lo1]} R. Longo ``A Duality for Hopf-Algebras''
 Comm. Math. Phys. {\bf 159} (1994) p.137
\bibitem[Lo2]{[Lo2]} R. Longo ``Index of Subfactors and Statistics
of Quantum Fields'' Comm. Math. Phys. {\bf 126} (1989) p. 217
\bibitem[Lo3]{[Lo3]} R. Longo ``Index of Subfactors and Statistics
of Quantum Fields II'' Comm. Math. Phys. {\bf 130} (1990) p. 285
\bibitem[Lo4]{[Lo4]} R. Longo``Solution of the factorial Stone-Weierstrass
Conjecture''
Invent. Math. {\bf 76} (1984) p. 145

\bibitem[N-W]{[N-W]}F. Nill, H.-W. Wiesbrock   in preparation
\bibitem[Oc1]{[Oc1]} A. Ocneanu ``A Galois Theory for Operator Algebras''
Notes of a Lecture
\bibitem[Oc2]{[Oc2]}  A. Ocneanu ``Quantum Symmetry, Differential Geometry
of Finite Graphs and Classification of Subfactors'' Lectures of A. Ocneanu
given at the University of Tokyo, Notes taken by Y. Kawahigashi (1992)
\bibitem[Pe]{[Pe]} G. Pedersen ``$C^*$-algebra and their automorphism groups''
Academic Press N.Y. (1979)
\bibitem[Pi-Po]{[Pi-Po]} M. Pimsner, S. Popa
``Entropy and Index for Subfactors''
Ann. Sci. Ec. Norm. Sup. {\bf 19} p. 57 (1986)
\bibitem[Po-Wo]{[Po-Wo]} Podles, S. Woronpwics ``Quantum deformations
of Lorentz group ''
Comm. Math. Phys.
{\bf 111} (1987) p. 613
\bibitem[St]{[St]} S. Str$\breve{a}$til$\breve{a}$ ``Modular Theory'' Abacus
Press (1981)
\bibitem[St-Zs]{[St-Zs]} S. Str$\breve{a}$til$\breve{a}$, L.Zsido ``Lectures on
von-Neumann Algebras'' Abacus Press (1975)
\bibitem[Szy]{[Szy]} W. Szymanski
 ``Finite Index Subfactors and Hopf Algebra Crossed Products",
Proceedings of the AMS Vol 120 No.2 (1994) p. 519
\bibitem[Ya1]{[Ya1]} S. Yamagami private communication
\bibitem[Ya2]{[Ya2]} S. Yamagami
``On Ocneanu's Characterization of Crossed Products''
Preprint (1993)
\bibitem[Ya3]{[Ya3]} S. Yamagami
``On Unitary Representation Theories of compact Quantum groups''
``Non Existence of Minimal Actions for Compact Quantum Groups''
Preprints (1993)
\end{thebibliography}
\end{document}